# Acetylene – Argon Plasmas Measured at a Biased Substrate Electrode for Diamond-Like Carbon Deposition, Part 1: Mass Spectroscopy.

A Baby, C M O Mahony and P D Maguire

Nanotechnology and Integrated Bio-Engineering Centre (NIBEC), University of Ulster, Newtownabbey, BT37 0QB, N. Ireland

**Abstract:** We report, for the first time, quadrupole mass spectrometry of neutral and positive ionic hydrocarbon species measured at the rf biased substrate electrode of an inductively coupled plasma for acetylene rich $C_2H_2$:Ar mixtures under various bias, frequency and pressure conditions. It has been observed that, irrespective of initial gas mixture, the resultant plasma is dominated by argon neutrals and ions. This is attributed to highly efficient conversion of acetylene to $C_2H$ due to the enhanced electron density compared to a standard capacitive plasma where the acetylene (neutral and ion) species remain dominant. This conversion may be crucial to film formation via inert rather than hydrocarbon ion bombardment. In addition, the transient formation of $CH_4$ from acetylene has been discovered using IR absorption spectroscopy with time constants similar to observed pressure variations. Rate coefficients and rates for many of the reaction mechanisms, calculated using measured EEDFs and species densities, are given. These results have important application in plasma models and growth studies for hydrogenated amorphous or diamond-like carbon film deposition. Film growth under similar plasma conditions is reported in an associated paper along with ion energy distributions for important growth species.

**Introduction**

Diamond-like or amorphous carbons (DLC), particularly hydrogenated amorphous carbon (a-C:H), represent a class of technologically important thin film materials. The ability to vary properties such as hardness, Young's modulus, surface roughness, electrical resistance, thermal conductivity, density, refractive index, offers considerable versatility in mechanical, electrical, optical and more recently biomedical applications [1]. These films are routinely deposited from hydrocarbon precursors, particularly $C_2H_2$, in conventional radio frequency (rf) capacitively-coupled plasma (CCP) systems [2-8], expanding thermal plasma [9, 10] and to a limited extent from inductively coupled plasma (ICP) systems [11-13]. These studies focussed mainly on material properties dependence on the key input variables, e.g. chamber pressure, power, flow rate etc. However future applications, for example in the biomedical arena, will require deposition at higher pressures or onto complex three-dimensional shapes. Hence a much greater understanding of film growth mechanisms which in turn depend on plasma – substrate conditions is necessary.

The growth of hydrogen-free DLC (tetragonal amorphous carbon, t-aC) is well understood in terms of carbon ion bombardment as the dominant species and sub-plantation or incorporation of the carbon into the growing film [14-16]. However hydrogenated DLC (a-C:H) has greater potential application, yet its growth mechanism is highly complex and as yet not understood [17]. A critical impediment is the lack of knowledge of substrate bombardment species, their energies and fluxes, from hydrocarbon-based plasmas, particularly acetylene, which is the preferred precursor gas over methane [4, 16]. There has been some attention given to gas phase species and reactions in methane containing plasmas [2, 11, 13, 18, 19] and to a much more limited extent to pure acetylene plasmas [20-23] for understanding polymerisation reactions, nanoparticle (dust) growth and DLC film formation. Vasile and Smolinsky [23] examined by mass spectrometry the ion chemistry of a pure and mixed acetylene discharges to determine the major reaction paths and highlight the main neutral-neutral and ion-neutral reactions. In their work on dusty plasmas, Deschenaux et al [20] studied mass and energy spectra of dominant ions and neutrals in methane, acetylene and ethylene plasmas. Macek et al [24] used mass-energy spectroscopy of $C_2H_2$/Ar in a triode ion plating apparatus and revealed a high degree of acetylene decomposition. However the spectra were complicated by the presence of Ti species from an in-situ evaporation source. Rangel et al [25] studied the electrical and optical properties of a polymerising $C_2H_2$/Ar plasma where OES was used to study mainly H and CH neutral species. Doyle et al [22] compared the kinetics of a methane rf glow discharge with measured gas species production and depletion rates in pure acetylene discharges. One dimensional fluid models for pure acetylene plasmas were developed by De Bleecker [21] and Herrebout D. [26] in order to understand the basic electron impact, ion-neutral and neutral-neutral reactions underlying the growth of nanosized (or dust) particles. Other models include a zero dimension kinetic model [27] of a rf $C_2H_2$ (1%)/$H_2$/Ar-rich plasma to determine the influence of pressure and power in the synthesis of nanodiamond thin films.

The existing experimental data and models of $C_2H_2$ based plasmas species have provided a framework for understanding the complex gas-phase reactions however their applicability to film growth conditions is much less appropriate. Typical deposition pressures are much lower (< 10 mTorr) than reported to date and high quality film growth normally requires $C_2H_2$ dilution with 30% - 90% inert gas (Ar). More importantly, the competing growth/etch mechanisms depend upon complex synergistic interactions between bombarding ions, radicals and the dynamic film surface [17]. Therefore, greater insight requires direct measurement of species flux and energy at the substrate under realistic growth conditions. Studies have been undertaken to measure substrate bombarding species using retarding field analysers, but mainly in etch gases [28-30]. However, mass resolution is not possible with this configuration. Mass and ion energy spectrometry has been reported for $CF_4$ plasmas [31] and for a $CH_4$ ECR source [32]. However, in neither case was the spectrometer located at the biased substrate electrode and thus the energies of growth/etch species could not be accurately determined.

In this paper we present ion and neutral species flux measurements obtained at the biased substrate under low pressure deposition conditions in a $C_2H_2$:Ar plasma. Two standard plasma systems were investigated, namely a capacitively-coupled plasma (CCP) and an inductively-coupled plasma (ICP) discharge. The former has been in routine use for the production of DLC films [1, 4-7, 16] and has been modified (increased chamber height) to accept a port for attachment of the mass-energy spectrometer to the chamber wall. The ICP system has been custom-designed to accommodate a range of optical and electrical diagnostics, including the incorporation of the mass-energy analyser into the rf-biased sample substrate. Additional diagnostics in the ICP system include a Langmuir probe, optical emission spectroscopy and quantum-cascade laser infra-red absorption spectroscopy for determination of the dominant neutral species in the bulk plasma. In a second companion paper [33], we present mass-resolved ion energy distribution spectra under a range of low pressure plasma conditions for direct comparison with deposited DLC film quality. Deposited film properties are evaluated under a range of conditions in order to derive guiding principles for process optimisation in terms of mean ion energy, ion energy spread and temporal neutral evolution and their relationship to bias voltage and frequency, pressure and gas flow ratio. In addition we also present some high pressure data for species flux and film growth.

The measurements of neutral and positive ionic species are the first presented for $C_2H_2$:Ar as the working gas. A complex network of reactions paths can be expected under our plasma conditions. However, we consider our observed species fluxes in terms of the simplest plausible reaction paths in order to highlight areas worthy of more detailed study. The formation of negative ions is also likely [20] but here we focus particularly on electropositive species and neutrals most favourable for DLC growth.

## 2. Experimental Details

The measurements described in this study were performed in two rf driven plasma facilities, one employing a Capacitively Coupled Plasma (CCP) and the other an Inductively Coupled Plasma (ICP). The working gases were usually pure argon or 2:1 $C_2H_2$:Ar mixtures; the small number of measurements made with other flow ratios are noted in the text.

### 2.1 CCP

The CCP comprised a 360 mm diameter 480 mm tall cylindrical chamber attached to associated gas, vacuum and electrical supplies. Gas flow controllers were used to set the flow rates (and thus ratios) of the working gases over the pressure range studied. A 13.56 MHz rf power supply and matching network was used to apply an rf potential to the 270 m diameter driven (substrate) electrode, sited centrally at the base of the chamber. To measure neutral fluxes, ion fluxes and energies, the chamber was adapted to house a HIDEN EQP 300 Mass Energy Analyser (MEA), which was sited at the side wall of the chamber, 170 mm above the top face of the driven electrode. The MEA was operated in three analysis modes: Neutral Mass (NM) for the mass resolved neutral flux, Ion Mass (IM) for mass resolved positive ion flux, and Ion Energy (IE) to give the Ion Energy Distribution (IED). These measurements were made for pressures 2 to 33 mTorr and negative CCP dc bias ($V_{dc}$) voltages of 0 V to 475 V. The upper limit of $V_{dc}$ represents an input power of ~ 130 W.

### 2.2 ICP

The ICP chamber (Figure 1) is a 400 mm diameter sphere with six orthogonal 250 mm diameter ports which provide access for plasma systems and diagnostics. Needle valves were used to set the flow rates and ratios of the working gases within a pressure range of 3.3 mTorr to 120 mTorr. The 13.56 MHz rf power supply and close coupled automatic matching network drives an rf current through the water cooled 152 mm diameter flat spiral ICP coil, which has four lobes to inhibit E-mode operation. The oscillating rf current in the coil couples to the plasma via a 12.5 mm thick quartz vacuum window below which a 120 mm diameter substrate electrode is sited. The window to electrode separation was 160 mm for all measurements. A second, variable frequency ($f_{bias}$), rf power supply and matching network was used to apply a negative bias ($V_{bias}$) to the substrate electrode. The substrate electrode was coupled, via a central 50 μm diameter hole, to the MEA, which was again operated in NM, IM and IE modes. Infrared absorption spectroscopy was used to measure line integrated species densities of $CH_4$ and $C_2H_2$ using a NeoPlas QCL-MACS system with a multi-pass White-cell arrangement [34]. The measurement axis crossed the vertical plasma axis 110 mm above the substrate electrode. A Scientific Systems compensated Langmuir probe was used to measure electron energy distribution functions (EEDFs), number densities, plasma and floating potentials 107 mm above the bias electrode, at a radius of 45 mm from the vertical axis. In house analysis was used to determine the component distributions of the measured EEDFs. A Scientific Systems Ion Flux Probe (IFP) was used to measure the total positive ion flux at the side of the plasma 50 mm above the bias electrode and 145 mm from the axis. This study concentrates on the abundance of neutral and positive ion species with varying pressure, $V_{bias}$ and $f_{bias}$. In most cases the ICP input power was set at 200 W and automatically matched so that the reflected power was < 1 W. The working gases were again pure argon or 2:1 $C_2H_2$:Ar.

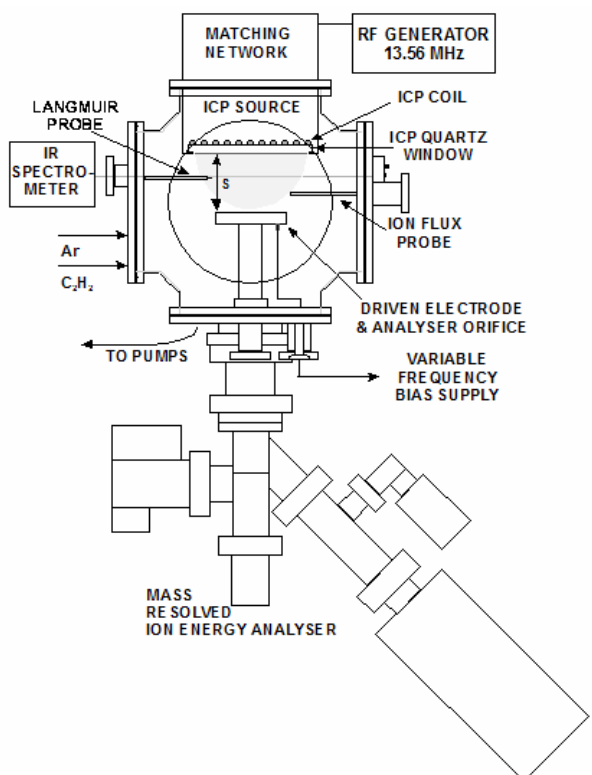

Figure 1: ICP chamber and diagnostics schematic diagram.

## 3. Results

We present data for both neutral and positive ion species in the CCP and ICP discharge systems described above. We do not consider negative ions because the negative ion flux from the plasma is assumed to be insignificant due to trapping in the positive plasma potential. Species identification is usually straightforward, however it is worth noting that although we assign $C_2H_4$ to mass 28 this peak is liable to masking by $N_2$ or CO. In addition, $C_2H_6$ has been assigned to mass 30 although a contribution by NO is also possible.

For all $C_2H_2$:Ar gas mixtures in both CCP and ICP systems, the set chamber pressure decreased after plasma turn on. For 475 V $V_{dc}$ CCP discharges with an initial $C_2H_2$:Ar flow ratio of 2:1, the pressure typically reduced to 73% of the set pressure. In the ICP discharges, for 200 W rf power, the pressure approaches the argon partial pressure, with typical time constants around 3 to 10 s, depending on set pressure. The pressure values quoted here are the equilibrium "plasma on" values: gas flow ratios, however, are those set beforehand i.e. for "plasma off" pressures.

### 3.1 CCP Neutrals

Figure 2 shows the count rate of a set of neutral species normalised to pressure and plotted as a function of pressure for a $C_2H_2$:Ar mixture (flow ratio 2:1) in the CCP system described in section 2. The MEA was mounted on the grounded chamber wall and operated in NM mode. We observe two distinct sets of species, one whose normalised count rate is constant with pressure; this includes atoms (Ar, C, H, $Ar^{2+}$), molecules ($C_2H_2$, $C_2H_4$, $C_4H_2$) and radicals ($C_2H$, $C_2$, $C_2H_3$, CH, $CH_3$ and $C_4H_4$), figure 2(a). The average ratio across this pressure range (shown in parenthesis) is calculated with respect to that of argon and shows the dominance of the ethnyl ($C_2H$) radical while the methyl ($CH_3$) radical was found to have almost the lowest concentration in this set, just above $C_4H_4$, the heaviest observed species. Other hydrocarbon species $C_xH_y$ (x = 1 to 4, y = 0 to 4) in addition to $Ar^{2+}$ were found to be of intermediate concentration. The second set consists of the remaining species, namely: $H_2O$, OH, $CO_2$ and ArH. Their normalised count rates decreased with increasing pressure, figure 2(b), with almost identical trends. The dominant species in this set, $H_2O$, has a similar concentration to $C_2H$ at low pressure.

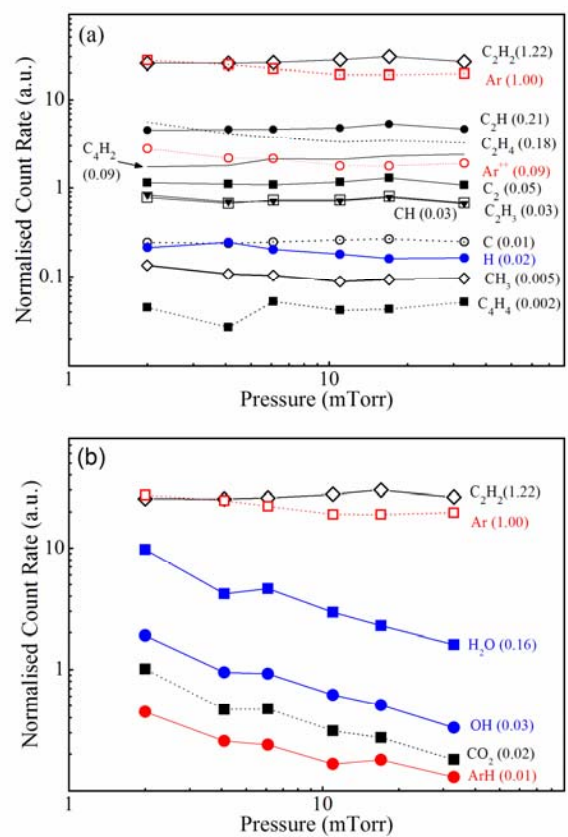

Figure 2 : Normalised count rate of neutral species in $C_2H_2$:Ar (flow ratio 2:1) CCP vs pressure at $V_{dc}$ = 450 V and f = 13.56 MHz . The ratios with respect to the argon value are shown in parentheses. (a) Species with little variation with pressure. (b) Species which fall with pressure.

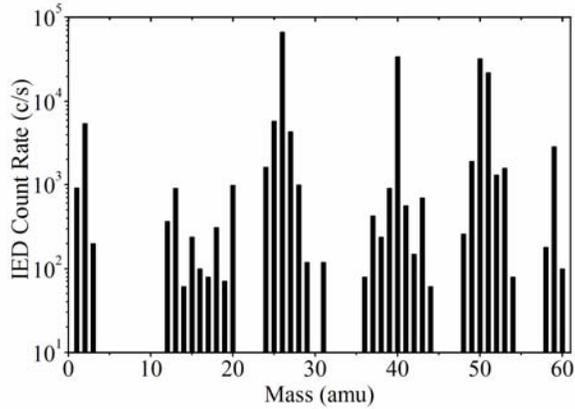

Figure 3: Ion mass spectrum rate in the CCP: 6 mTorr $C_2H_2$:Ar (flow ratio 2:1) at $V_{dc}$ = 450 V and f = 13.56 MHz. The ion energy was set at 30 eV to avoid saturation.

**3.2 CCP Positive Ions**

Positive ion count rates were obtained in IM mode with the spectrometer attached to the grounded chamber wall and scanned over the mass range 0 – 60 amu at an ion energy of 30 eV, chosen so as to prevent count saturation. Figure 3 shows a typical contribution of positive ions to the mass flux for a fixed pressure (6 mTorr), $C_2H_2$:Ar mixture (flow ratio 2:1) and 450 V substrate bias. The dominant ion species are $C_2H_2^+$, followed by $Ar^+$, $C_4H_2^+$ and $C_4H_3^+$. Other significant species include $C_2H^+$, $H_2^+$, $H_3^+$ $C_2^+$, $C_2H_3^+$ and $C_2H_4^+$. Among the heavier mass ions, $C_4H^+$, $C_4H_4^+$ and $C_4H_5^+$ are also abundant.

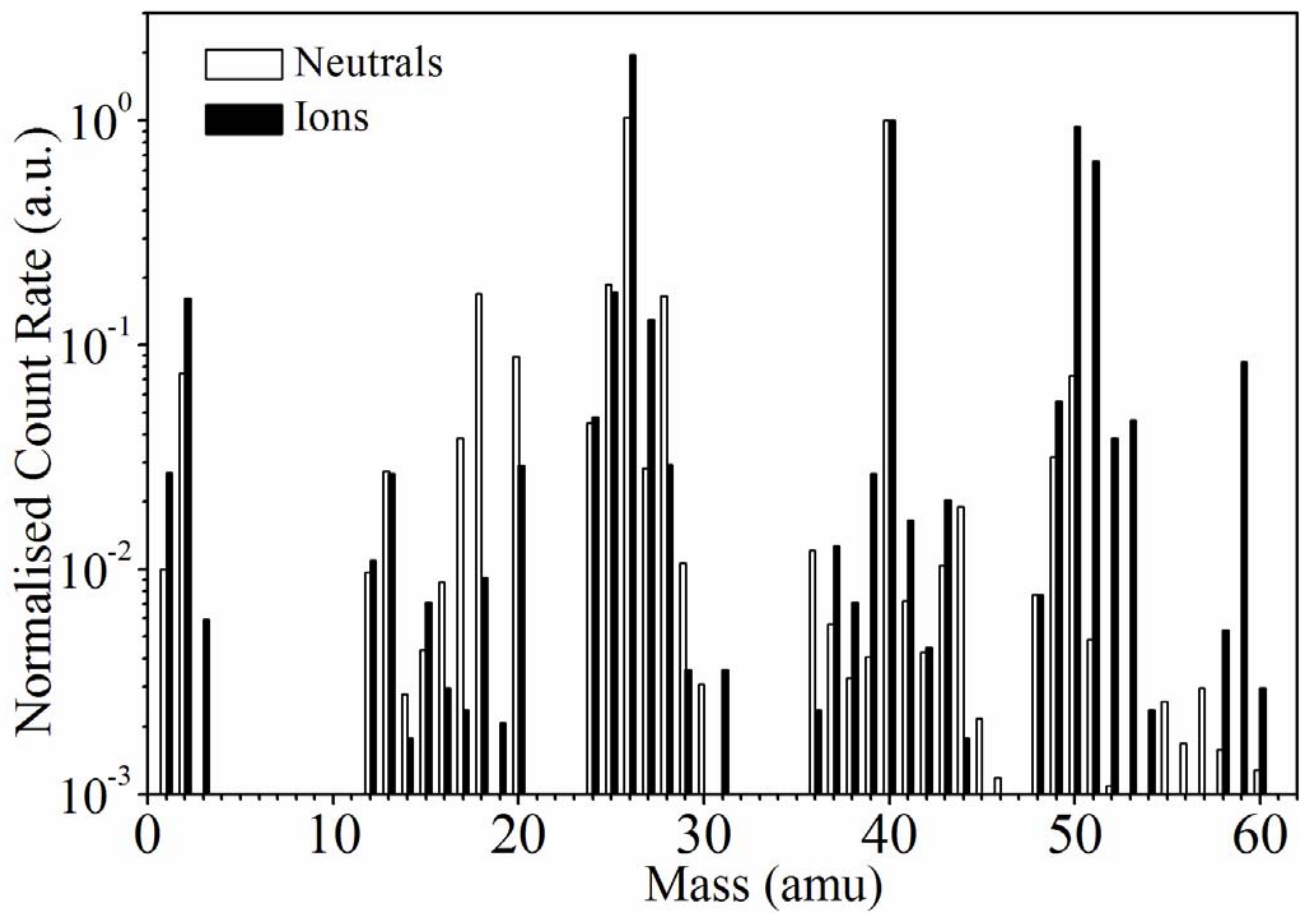

Figure 4: Neutral and ion mass spectra normalised with respect to argon in the CCP: 6 mTorr $C_2H_2$:Ar (flow ratio 2:1) at $V_{dc}$ = 450 V and f = 13.56 MHz. The ion energy was set at 30 eV to avoid saturation.

Since direct quantitative comparison of neutral and ion count rates is not possible due to the different sensitivities in NM and IM modes, we compare relative fluxes by first normalising them to their respective argon count rate, figure 4. For some species the normalised neutral count is greater than that of its ion: OH, $H_2O$, $Ar^{2+}$, $C_2H_4$, $C_3$ and $CO_2$. For others the converse is the case: $H^+$, $H_2^+$, $C_2H_2^+$, $C_2H_3^+$, $C_3H^+$, $C_3H_3^+$, $ArH^+$, $C_4H^+$, $C_4H_2^+$, $C_4H_3^+$, $C_4H_4^+$, $C_4H_5^+$. Similar normalised ion and neutral counts are observed for C, CH, $C_2$ and $C_2H$. Acetylene ions are the only species which shows a ratio > 1.

The effect of varying plasma power and thus $V_{dc}$ on neutral count rates is shown in figures 5. The initial pressure with no plasma ($V_{dc}$ = 0 V) was 8.6 mTorr; this dropped almost linearly with increasing bias to 6.2 mTorr at $V_{dc}$ = 475 V. The numbers in parenthesis after the species legends are the ratios of counts at a $V_{dc}$ of 475 V, normalised to that of argon. The bias value chosen represents optimum bias condition for film deposition.

The neutral species can be divided into three sets. Group I consists of the carbon atom and some carbon bearing radicals (C, $C_2$, $C_2H$, $C_2H_3$ and CH) which directly follow the exponential reduction of $C_2H_2$ with increasing $V_{dc}$, figure 5(a). This reduction in count rate with bias is about 3 times greater than the drop in pressure noted above. In figure 5(b) Group II comprises Ar, H & $CH_3$, which are almost bias independent. Group III (also figure 5(b)) consists of other molecules ($H_2O$, $CO_2$, $C_4H_2$) and radicals (ArH, OH $C_4H_4$, $C_2H_4$) which display more complex behaviour with bias. For example, $C_4H_2$ and $C_4H_4$ count rates increase significantly at plasma turn on ($V_{dc}$ > 0 V). Further, in all groups, all observed species except $C_4H_2$ and $C_4H_4$ are present in significant quantities when the plasma is off ($V_{dc}$ = 0 V).

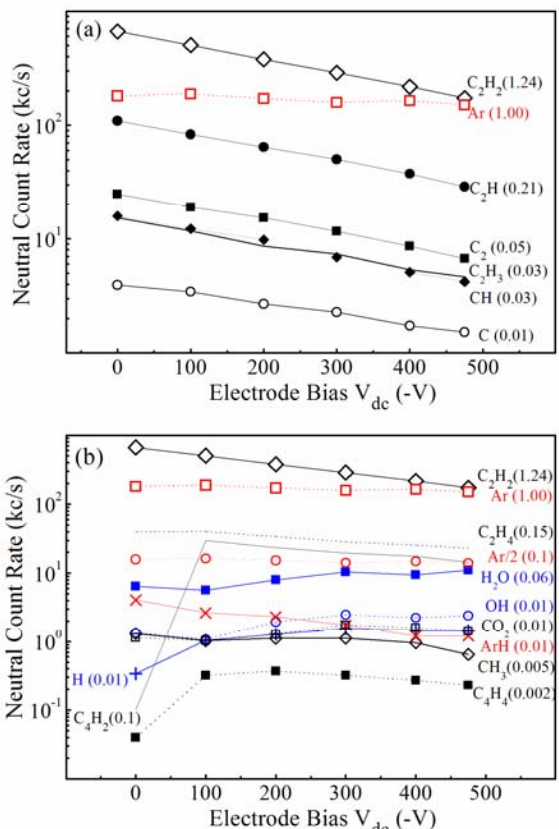

Figure 5: Neutral species variation with substrate bias in the CCP: starting pressure = 8.6 mTorr $C_2H_2$:Ar (flow ratio 2:1) and f = 13.56 MHz.
The pressure dropped from 8.6 mTorr at $V_{dc}$ = 0 V to 6.2 mTorr at $V_{dc}$ = 475 V. The ratios with respect to the argon value are shown in parentheses. (a) Species which reduce with $V_{dc}$ (Group I), (b) Species which are almost constant with $V_{dc}$ (Group II) and species with more complex variation with $V_{dc}$.(Group III). For comparison, the argon and acetylene data are shown in both (a) and (b).

Since the spectrometer is positioned at the grounded wall we cannot directly determine the species flux to the driven electrode. However the open geometry of this CCP means that the ratio of wall flux will be similar to those at the driven electrode.

Figure 6(a) shows the ratio of count rates normalised to that of $C_2H_2$ at $V_{dc}$ = 475 V for group I members. The dominant $C_2H$ radical is an important precursor for a-C:H growth, (Figure 6(a)). A similar plot for group II and III, figure 6(b), shows the significant hydrocarbon species to be $C_2H_4$ and $C_4H_2$.

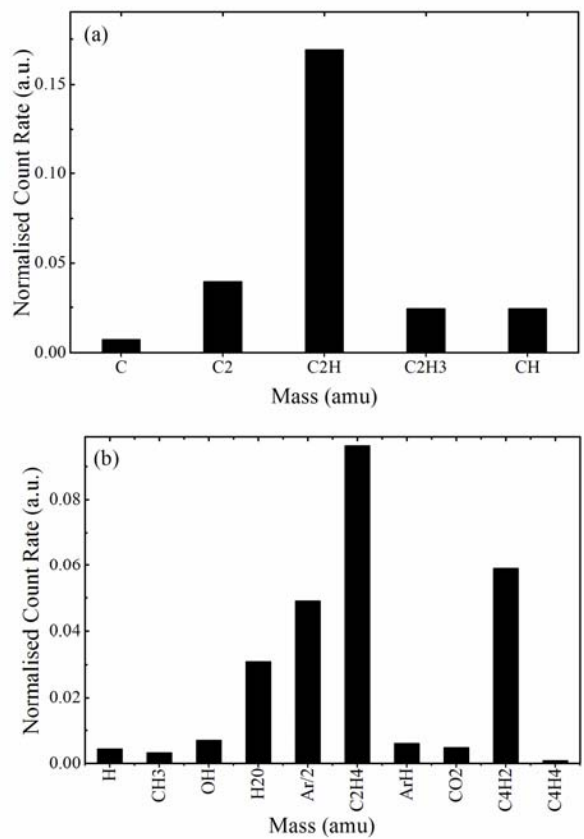

Figure 6: Neutral species normalised to $C_2H_2$ at $V_{dc}$ = 475 V in the CCP: 6.2 mTorr $C_2H_2$:Ar (flow ratio 2:1) and f = 13.56 MHz. (a) Group I members. (b) Groups II and III.

### 3.3 ICP neutrals

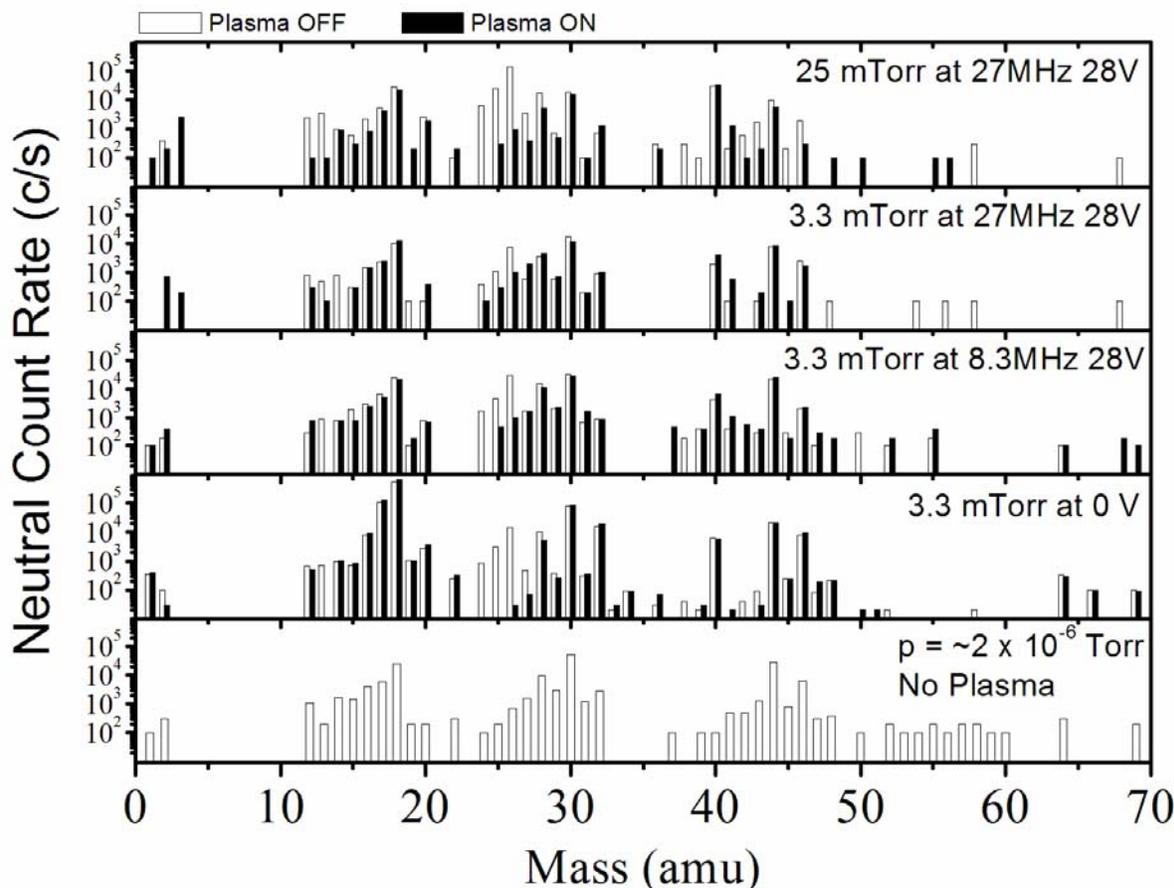

Figure 7: Neutral mass spectra in the ICP $C_2H_2$:Ar (flow ratio 2:1) for various pressures, $V_{bias}$ and $f_{bias}$. Data in black shows plasma on. Data in white shows plasma off. Also included is a mass spectrum at base pressure $\sim 2 \times 10^{-6}$ Torr.

Figure 7 shows examples of the neutral count rates measured for both plasma off and plasma on for a range of pressures and electrode bias conditions. For all data the $C_2H_2$:Ar flow ratio was 2:1 and ICP power was 200 W. The upper scan limit was set at 70 amu to maximise analyser operational lifetime. All data were recorded under the same MEA measurement conditions, so count rates are directly comparable. Neutral count rates at a base pressure of $2 \times 10^{-6}$ Torr (plasma off) show the nature and level of background impurities, namely: water vapour $H_2O$, $N_2$, and $C_2H_6$ (or NO).

Plasma ignition resulted in changes in the species concentrations as well as the production of new species or destruction of old species before (white) and after plasma turn on (black). $H_3$, $C_4$, $C_4H_2$, $C_4H_7$ and $C_4H_8$ were observed only post-ignition at high bias frequency (27 MHz). $C_5H_8$ and $C_5H_9$ were the only new species observed at low bias frequency. New $H_n$ species (n = 1 – 3) were observed, mainly at high pressure. Though the production of $H_3$ is observed only at high bias frequency, its count rate at high pressure was 10 times greater than at low pressure. Significant reduction in C and CH concentrations were observed as pressure was raised from 3.3 mTorr to 25 mTorr. $C_2$ is only observed in the plasma off state. Argon neutral count remained constant while ArH variation was much more significant. Under all plasma conditions a significant decrease in $C_2H_2$ and $C_2H$ concentrations, the dominant carbon-carrying species, was observed compared to the plasma off state.

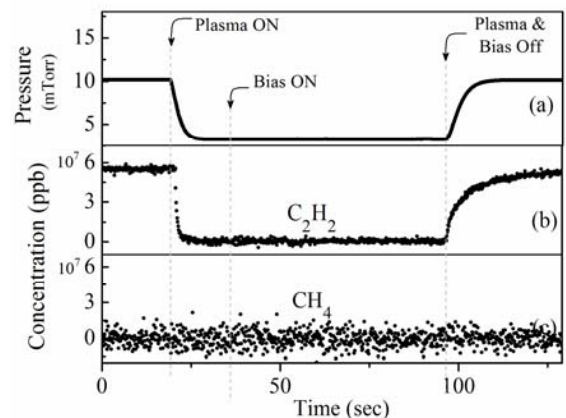

Figure 8: Chamber pressure and $C_2H_2$ & $CH_4$ concentrations vs time in the ICP: 10 mTorr set pressure, $C_2H_2$:Ar (flow ratio 2:1). The plasma was turned on at t=20 s, and bias (28 V, 8.311MHz) at 35 s. Both plasma and bias was turned off at 100 s.

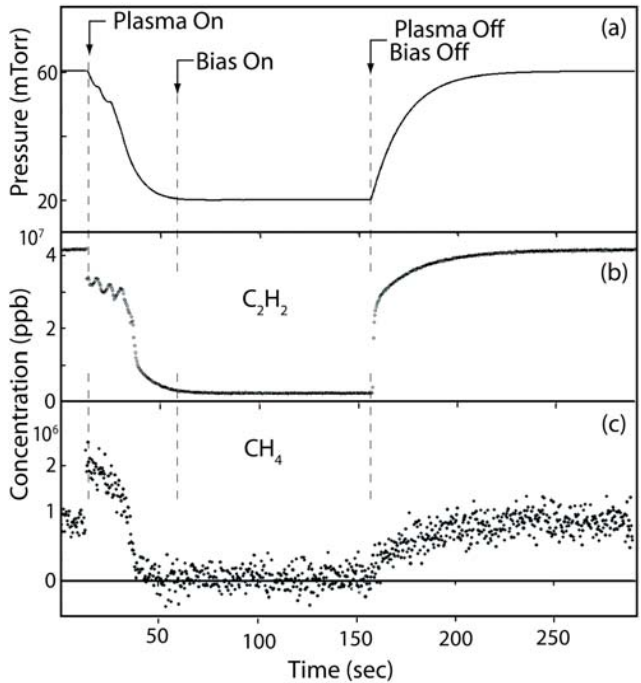

Figure 9: Chamber pressure and $C_2H_2$ & $CH_4$ concentrations vs time in the ICP: 60 mTorr set pressure, $C_2H_2$:Ar (flow ratio 2:1). The plasma was turned on at t ~15 s, and bias (28 V, 8.311MHz) at 60 s. Both plasma and bias was turned off at ~ 160 s.

The fall in pressure on plasma ignition and the loss of $C_2H_2$ was investigated in more detail by following the time evolution of the pressure and the IR absorption signal for $C_2H_2$ and $CH_4$ species upon plasma switch-on, figure 8 (10 mTorr). The IR absorption measurements provide line integrated concentration values in the bulk plasma 110 mm above the substrate. At 10 mTorr, the $C_2H_2$ signal is seen to fall with an identical time constant τ to that of the pressure (2.5 s). The $CH_4$ signal is barely visible above the background noise, however at higher pressure, figure 9, its presence is significant. Here the $CH_4$ is seen to rise rapidly (< 1 s) upon plasma ignition and then fall with a similar time constant to that of pressure and $C_2H_2$ (~20 s), indicating $CH_4$ as a transient by-product of $C_2H_2$ dissociation. It can be seen (figures 8 & 9) that bias has no effect on pressure or these species concentrations.

In a separate experiment, a compensated Langmuir probe [35] 107 mm above the bias electrode was used to measure changes to bulk plasma parameters as acetylene was added to an argon discharge. Parameters were recorded every 5 s as a 3.3 mTorr 200 W Ar plasma was transformed by the introduction of $C_2H_2$ at T~ 40 s. This addition of 16 sccm $C_2H_2$, twice the Ar flow rate, caused the chamber pressure to rise to 4.1 mTorr. The $C_2H_2$ flow was turned off at T~ 170 s. The addition of acetylene to the plasma volume led to significant changes in the plasma, figure 10. The values of $V_{pl}$, $T_e$, and $n_e$ dropped within 5 s (cf τ ~ < 2.5 s figure 8) to near constant levels: $n_e$ by a factor of 2, $T_e$ by ~5% and $V_{pl}$ from 22 V to 20 V. For a short period (120 s < T < 145 s) the sample electrode bias was turned on so that $V_{bias}$ = 100 V; no significant changes were observed. $V_{pl}$ and $T_e$ were restored to their original values within 30 s of $C_2H_2$ turn off, whereas the $n_e$ recovery was slower. Note that for this study, $T_e$ was determined simply from the slope of the compensated Langmuir probe IV characteristic.

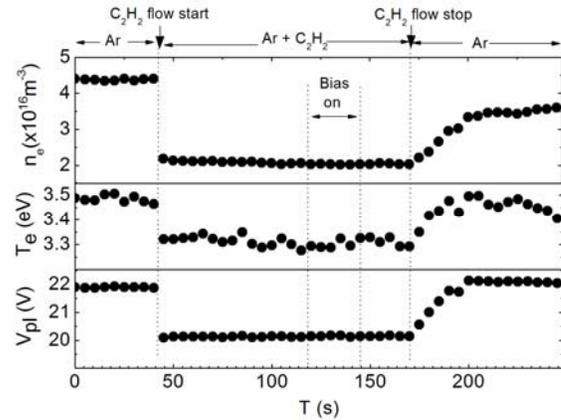

Figure 10: Plasma potential ($V_{pl}$), electron temperature ($T_e$) and electron density ($n_e$) vs time in the ICP. The 200 W plasma was ignited in pure argon (3.3 mTorr) at T < 0s and $C_2H_2$ was added at T=40s (with a $C_2H_2$:Ar *flow* ratio of 2:1). Bias (28 V, 8.311MHz) was applied for 120s < T < 145 s. The $C_2H_2$ flow was turned off at T 170 s.

### 3.4 ICP Positive Ions

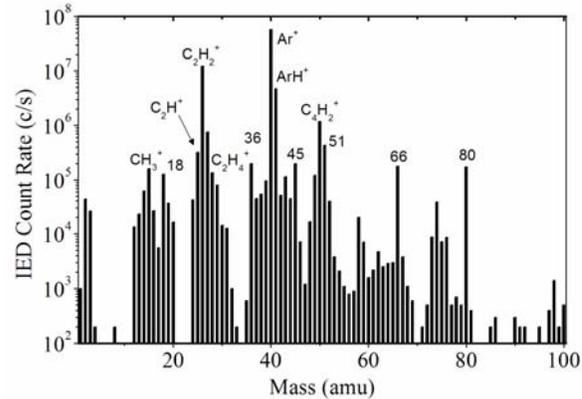

Figure 11: Ion mass spectrum in the ICP: 4.5 mTorr $C_2H_2$:Ar (flow ratio 1:20), 40 W 13.56 MHz and electrode grounded.

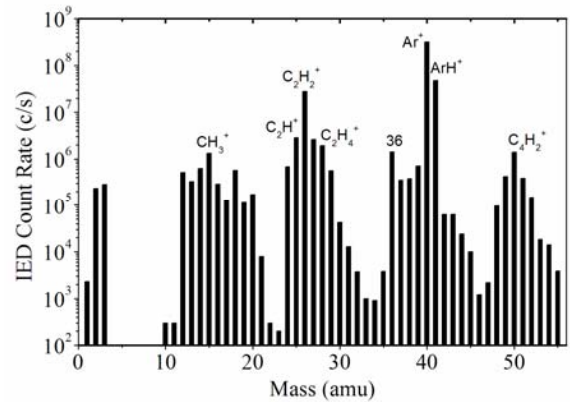

Figure 12: Ion mass spectrum in the ICP: 3.3 mTorr $C_2H_2$:Ar (flow ratio 2:1), 200 W 13.56 MHz and substrate bias 7 V @ 8.311 MHz.

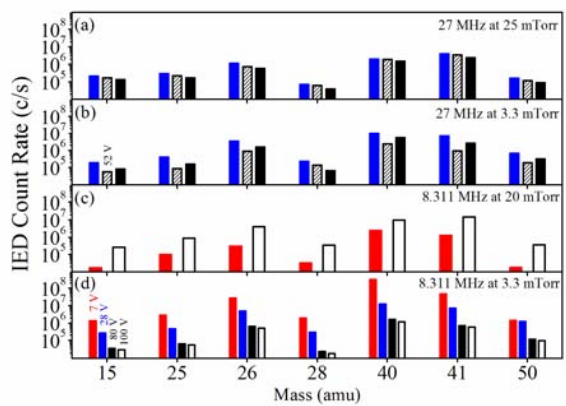

Figure 13: Dominant ionic species ($CH_3^+$, $C_2H^+$, $C_2H_2^+$, $C_2H_4^+$, $Ar^+$, $ArH^+$, $C_4H_2^+$) in the ICP: $C_2H_2$:Ar (flow ratio 2:1), 200 W 13.56 MHz for various pressures, $V_{bias}$ and $f_{bias}$.

Figure 11 shows positive ionic species up to mass 100 amu for a 40 W E-mode ICP discharge with a $C_2H_2$:Ar gas flow ratio of 1:20 and pressure of 4.5 mTorr. Ions from the two feed gases $Ar^+$ and $C_2H_2^+$ dominate, and with $ArH^+$ and $C_4H_2^+$, make up 95% of the total count. The majority of species with count rates greater than $10^3$ times that of $Ar^+$ are identified as hydrocarbons, and may be grouped into families based on carbon content:

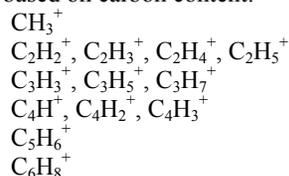

$CH_3^+$
$C_2H_2^+$, $C_2H_3^+$, $C_2H_4^+$, $C_2H_5^+$
$C_3H_3^+$, $C_3H_5^+$, $C_3H_7^+$
$C_4H^+$, $C_4H_2^+$, $C_4H_3^+$
$C_5H_6^+$
$C_6H_8^+$

Water and argon isotopes with their associated hydrides make up the remainder.

Similar data for the higher power H mode ICP discharge is shown in figure 12 for a 200 W plasma with a $C_2H_2$:Ar gas flow ratio of 2:1 and pressure of 3.3 mTorr. By comparison to the low power ICP case, $Ar^+$, $ArH^+$ and $C_2H_2^+$ contribute, in order of dominance, 95 % of the total count. The remainder (to 99.5 %) consists of $CH_3^+$, $C_2H^+$, $C_2H_3^+$, $C_2H_4^+$, mass 36 ($C_3^+$ & $^{36}Ar^+$), $C_4H_2^+$.

The count rates of the dominant ions ($Ar^+$, $ArH^+$, $C_2H_2^+$) plus those of possible importance in diamond like carbon growth ($CH_3^+$, $C_2H^+$, $C_2H_4^+$, $C_4H_2^+$) were measured for various pressure, bias voltage and bias frequency conditions in ICP H-mode (200 W) with a gas mixture $C_2H_2$:Ar flow flow ratio of 2:1, figure 13. The species count rates generally retain their order of abundance across the whole data set. $Ar^+$ or $ArH^+$ are either $1^{st}$ or $2^{nd}$, followed by $C_2H_2^+$ and then $C_2H^+$, $C_4H_2^+$ or $CH_3^+$ are either $5^{th}$ or $6^{th}$ and $C_2H_4^+$ usually has the lowest count rate. Measured count rates do, however, vary with pressure, bias voltage and frequency. Figure 13(c) and 13(d) show ion count rates for the above species at 8.311 MHz bias. The flux decreases consistently as bias is increased from 7 to 100 V at 3.3 mTorr, whereas the opposite behaviour is observed for the two biases recorded at 20 mTorr. At 27 MHz bias (figure 13(a) and 13(b)), the flux minimises at 52 V bias at 3.3 mTorr, whereas it decreases with increasing bias at 25 mTorr.

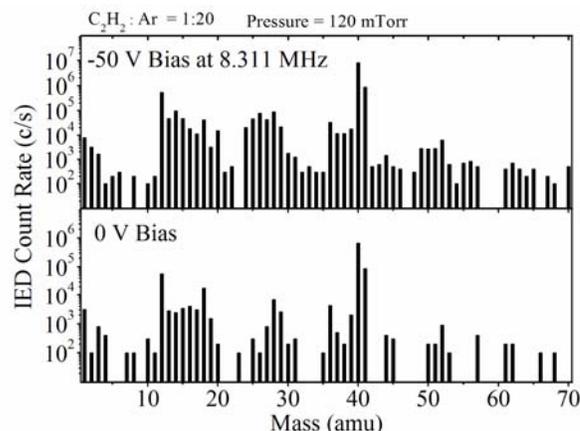

Figure 14: Ion mass spectrum in the ICP: 120 mTorr $C_2H_2$:Ar (flow ratio 1:20), 200 W 13.56 MHz for $V_{bias}$ = 0 V (electrode grounded) and $V_{bias}$ = 50 V bias and $f_{bias}$ = 8.311 MHz.

Figure 14 shows ion count rates measured at high pressure. A $C_2H_2$:Ar flow ratio of 2:1 was found not to be reproducibly stable and therefore a flow ratio of 1:20 was used. Positive ionic species count up to mass 70 amu were measured at 200 W ICP power for both biased (8.311 MHz, $V_{bias}$ = 50 V) and grounded electrode conditions. Ions $Ar^+$, $ArH^+$ and $C^+$ contributed ~92% of the total ions irrespective of the substrate bias. $C_2H_2^+$ which appears to be the dominant ion at $C_2H_2$:Ar flow ratio of 2:1 was found to contribute almost nil at grounded electrode whereas ~1% was observed at 50 V substrate bias. Other positive ions which were only produced at high substrate bias were $C_2H^+$, $C_2H_5^+$, $C_2^+$, $Ar^{++}$, $C_3H^+$, $C_3H_2^+$. In addition $CH_2^+$, $CH^+$, $CH_3^+$, $H_2O^+$, $C_3^+$, $CH_4^+$, $C_2H_3^+$, $C_3H_3^+$ were observed at both low and high biases whereas $H^+$ and $C_4H_4^+$ were observed only at 0 V bias.

### 4. Discussion

The low pressure acetylene-argon plasma is an important and widely used process in materials deposition but detailed plasma characterisation is lacking, not least because of the challenges of diagnostics in high rate deposition environments. The neutral and positive ionic species measured here in low pressure (3 – 120 mTorr) CCP and ICP systems are the first such measurements published with $C_2H_2$:Ar as the working gas. It should be noted that $C_xH_y$ based plasmas offer a wide range of complex reaction paths [9, 23, 27, 36-38]. This is further complicated by adding argon, so accurate determination of $C_2H_2$:Ar reaction probabilities requires additional detailed modelling.

In this paper we suggest mechanisms to explain some of the observations; the task is, however, too complex to account for them all. We have therefore considered our observed species fluxes in terms of the simplest plausible reaction paths in order to highlight areas worthy of more detailed study. In particular, the most important growth precursor for diamond-like carbon is the $C_2H$ radical [22]. Accurate measurement of this radical is particularly challenging and hence it is worth considering possible $C_2H_2$ to C$_2$H conversion mechanisms. Also of interest is the impact on growth models when the presumed dominance of C$_2$H$_2^+$ is replaced by Ar$^+$ or ArH$^+$, hence affecting the direct carbon ion subplantation rates.

The C$_2$H$_2$ partial pressure drop at plasma turn on for mixtures, typically from 6.7 mTorr to < 0.3 mTorr in the ICP, was noted in section 3, and is associated with a dramatic loss of the C$_2$H$_2$, as evidenced by IR absorption v time, figures 8 & 9. The typical argon partial pressure, 3.3 mTorr remained unaffected. Thus, the dominance of both Ar neutrals and ions, despite the input gas flow ratio, is unsurprising. In the CCP, it was observed that loss of C$_2$H$_2$ was dependent on the input power, figure 5. For powers up to 130 W (V$_{dc}$ = 475 V) C$_2$H$_2^+$ remains the dominant ion, an indication of the lower dissociation in the CCP versus the ICP, given the lower power coupling efficiency of the former.

One potential mechanism for C$_2$H$_2$ reduction is the hydrogen abstraction from acetylene by electron dissociation [38, 39] equation (1) and equation (2):

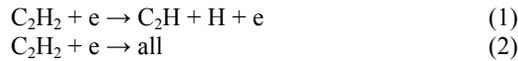

$$C_2H_2 + e \rightarrow C_2H + H + e \quad (1)$$
$$C_2H_2 + e \rightarrow \text{all} \quad (2)$$

This reaction would need to destroy C$_2$H$_2$ and produce C$_2$H at a rate equivalent to the ICP C$_2$H$_2$ gas flow rate (typically 24 sccm into the 36 litre chamber, ie ~ 3 x 10$^{15}$ m$^{-3}$s$^{-1}$). Indeed this is easily surpassed by the calculated ICP rate > 10$^{19}$ m$^{-3}$s$^{-1}$, (Table 2, Section 4.5).

Due to the high surface loss probability (β ~ 0.9) of the C$_2$H radical [40], the C$_2$H produced via equation (1) readily sticks to chamber wall surfaces and hence is lost from the volume. This conversion of neutral C$_2$H$_2$ combined with high surface loss probability of the C$_2$H radical may explain the initial drop in pressure. In effect the chamber walls are acting as an additional pumping source.

To our knowledge there are no other reports of this pressure phenomenon in studies of C$_2$H$_2$ with argon dilution at low pressure. This is most likely due to the dominance of CCP systems where the effect is not so noticeable. Our pressure control systems are operated in free-running mode where the total flow and flow ratio rather than pressure is set. Again this is standard practice. Nevertheless, those systems operated in automatic constant pressure mode should respond with considerable increases in flow and hence much greater deposition throughout. For pure C$_2$H$_2$ studies, Vasile and Smolinsky [23] report severe experimental difficulties with pure C$_2$H$_2$ due to rapid deposition of polymer-like material which blocks the spectrometer orifice and they needed to dilute the acetylene with rare gases.

Our suggested mechanism of C$_2$H$_2$ loss, equation (1), requires a decisive conversion to C$_2$H that is not however detected by the MEA. This may be due to the transmission function of the MEA orifice, which has an aspect ratio (height/diameter) of ~10. Thus, only C$_2$H molecules that arrive almost on-axis (within a solid angle of 0.04 steradian) will traverse the orifice without interaction with the wall. Due to the high sticking coefficient, most C$_2$H molecules with higher angles of incidence will stick to the wall, whereas all C$_2$H$_2$ molecules will be transmitted, even if wall contact is made. Thus the orifice, due to its geometry, can be expected to function as a blocking filter for C$_2$H. An estimate of the relative C$_2$H/C$_2$H$_2$ transmission probability is 7%. Note, that in order to prevent build-up of deposit on the orifice wall, thus reducing the transmission further, we undertake an Ar/O$_2$ plasma clean between each run. Without this step we can observe total orifice closure with deposited material.

Another factor that needs consideration is the transient production of CH$_4$ during the pressure drop at plasma turn-on, figure 9, as determined by IR absorption. At 10 mTorr the overall CH$_4$ concentration is below our detection level (in the presence of C$_2$H$_2$). However at higher pressure we note an immediate rise in CH$_4$ to 3.6% of the set C$_2$H$_2$ concentration, followed by a similar decay to that of pressure and C$_2$H$_2$, implying a direct C$_2$H$_2$-CH$_4$ relationship. Serdioutchenko et al [41] has also observed transient effects in a CH$_4$ – Ar plasma with CH$_4$ conversion to C$_2$H$_2$ and higher hydrocarbons. However the "reverse" conversion of C$_2$H$_2$ to CH$_4$ has not been observed and is not considered probable in the C$_2$H$_2$ literature. This is possibly due to the lack of time-resolved measurements of C$_2$H$_2$ to date. We can postulate a number of mechanisms and here we offer the simplest. Note that the transient evolution of species cannot be detected by the MEA since the measurement (and software synchronisation) time is approximately 3 s.

With sufficient H$_2$ concentration, the generation of CH$_4$ is possible from the hydrogenation of CH$_3$ [42], equation (3), which may in turn be generated from the reaction of the CH$_2$ radical with H$_2$ [42], equation (4).

$$CH_3 + H_2 \rightarrow CH_4 + H \quad (3)$$
$$CH_2 + H_2 \rightarrow CH_3 + H \quad (4)$$

Addition of argon may further aid in the formation of CH$_4$ [27], via catalytic action, most likely taking place at the wall.

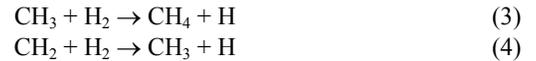
$$Ar + CH_3 + H \rightarrow CH_4 + Ar \quad (5)$$

the ionic versions are also possible

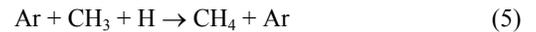
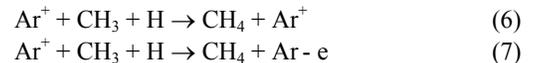
$$Ar^+ + CH_3 + H \rightarrow CH_4 + Ar^+ \quad (6)$$
$$Ar^+ + CH_3 + H \rightarrow CH_4 + Ar - e \quad (7)$$

This catalytic argon effect may explain the lack of observed CH$_4$ formation in pure C$_2$H$_2$ plasmas. The loss mechanisms for CH$_4$ are well documented [2, 11, 13]. As noted above C$_2$H is the dominant derived product of C$_2$H$_2$ dissociation, equation (1), within the plasma and a number of parallel reactions also exist, equation (2).

In the plasma off state, apart from input gases and impurity species, we have also observed other species such as C$_2$H, CH$_3$ etc. These may be indicators of derivative species generation within the MEA (in NM mode) due to electron dissociative ionisation of C$_2$H$_2$ in the ionisation source. The filament electron energy is normally set to 70 eV, about 5-15 eV above the energy at which maximum total ionisation cross section occurs [43]. As an example, we consider two reaction pathways within the MEA, namely:

$$e \text{ (filament)} + C_2H_2 \rightarrow C_2H_2^+ + 2e \quad (8)$$
$$e \text{ (filament)} + C_2H \rightarrow C_2H^+ + 2e \quad (9)$$

The first reaction, equation (8), will add to the $C_2H_2$ counts at the mass energy analyzer detector, while the second, equation (9), will add erroneous counts for the $C_2H$ radical and underestimate the $C_2H_2$ count. The measured ratio of $C_2H_2/C_2H$ without plasma, figure 15, is compared with the expected count ratio determined from calculated cross-sections for equation (8) and equation (9). The close fit confirms that the measured plasma-off $C_2H$ counts represent a derivative product created within the MEA rather than $C_2H$ from within the chamber. Thus the use of MEA for $C_2H$ detection has been shown to be highly problematic due to orifice, sticking and filament ionization effects.

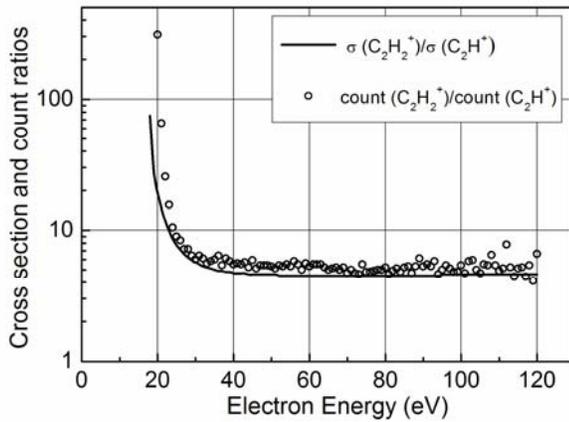

Figure 15: Measured $C_2H_2/C_2H$ count ratios vs MEA electron ionizer energy in the ICP: 10 mTorr $C_2H_2$:Ar (flow ratio 2:1), and plasma off. The line shows the ratio calculated from the cross sections for equations (8) & (9).

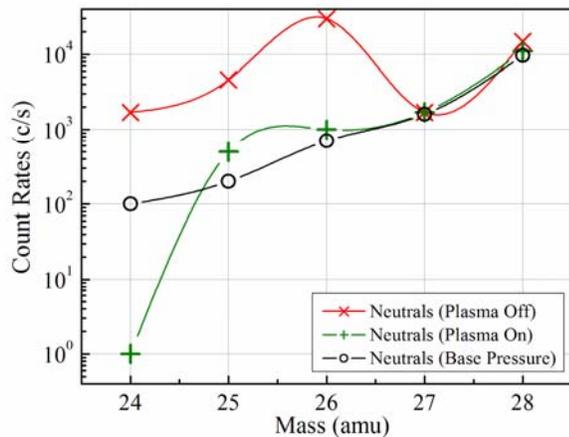

Figure 16: Neutral mass spectra around Mass 26. Green: ICP plasma on: 3.3 mTorr $C_2H_2$:Ar (flow ratio 2:1), 200 W 13.56 MHz, electrode grounded. Red: ICP plasma off: 10 mTorr $C_2H_2$:Ar (flow ratio 2:1) and Black: base pressure ($2 \times 10^{-6}$ Torr).

This raises the question as to whether any $C_2H$ generation from $C_2H_2$ occurs within the plasma volume and, if so, whether it is sufficient to account for the observed fall in pressure. Figure 16 shows neutral count rates, measured at electron energy of 70 eV, for masses 24 to 28, at base pressure ($10^{-6}$ Torr) and at 10 mTorr, with and without plasma. At 10 mTorr, with the plasma off, we assume all neutrals except $C_2H_2$ are caused by ionisation within the MEA. At plasma turn on $C_2H_2$ and $C_2H$ neutral fractions drop; $C_2H_2$ by a factor of 30 but $C_2H$ by only a factor of 10, indicating detection of some of the $C_2H$ formation within plasma. We estimate the film growth rate on the chamber walls due $C_2H$ deposition, assuming 100% conversion from $C_2H_2$, to be approximately 2 nm/s, in good agreement with measured substrate film growth rates of 1 nm/s [33]. Note that ion species detection in MEA IE mode is not subject to this effect since the electron filament is turned off for these measurements.

### 4.1 CCP neutral species

The neutral species variation with pressure has been classified into two families: the first, with densities approximately proportional to pressure, appear as near horizontal lines in figure 2(a). The second family, figure 2(b), shows a reduction of normalised count rate with increasing pressure, however this fall is not as steep as one would get from a species with constant count at all pressures.

We would expect the input gases to be in the first family, but the behaviour of the other species in this family is non-trivial. Excluding recombination, a first order approximation from particle balance gives the count rates normalised to pressure:

$$\frac{n_c}{n_0} = \frac{n_e n_p k}{n_0 L} \quad (10)$$

The rate coefficient for the particular reaction is k and L is an assumed constant loss factor; $n_0$, $n_e$, $n_p$ and $n_c$ are total neutral, electron, parent and child species number densities respectively. For first generation products, $n_p \propto n_0$, so we deduce that the product $n_e k$ is almost constant for these species across the pressure range investigated. The second family: $H_2O$, $OH$, $CO_2$ and $ArH$ in figure 2(b), are almost certainly due to impurities, and the variation with pressure, figure 2(b), suggest they are released from the chamber walls rather than being present in the $C_2H_2$ feedstock or from leaks.

Macek et al [24] used mass spectroscopy to measure species abundances and postulated relevant reactions. They observed atomic hydrogen to be the dominant neutral whereas in our case it is $C_2H_2$ followed by Ar. However, Macek's triode ion plating system contained significant Ti evaporant species used for the deposition of TiC in Ar + $C_2H_2$ working gas which may reduce the $C_2H_2$ abundance. In our study, the most abundant child molecule is $C_2H_4$, (Figures 2(a) and 5(b)) which can be formed by the neutral-neutral reaction [27, 43, 44] equation (11).

$$CH_2 + CH_3 \rightarrow C_2H_4 + H \quad (11)$$

with a rate coefficient $7.0 \times 10^{-17} \, m^3 s^{-1}$ [44] or by the electron collisional dissociation of $C_2H_6$ [39] equation (12).

$$e + C_2H_6 \rightarrow C_2H_4 + H_2 + e \qquad (12)$$

Neither $CH_2$, $CH_3$ nor $C_2H_6$ are observed to be particularly abundant in our CCP and other reactions may also need to be considered.

Deschenaux's [20] experimental observations, in pure $C_2H_2$ CCP, has shown a similar spectrum of major neutrals namely: $H_2$, CH, $C_2$, $C_2H$, $C_2H_2$, $C_2H_3$, $C_2H_4$, $C_4$, $C_4H$, $C_4H_2$ and $C_4H_3$ out of which $H_2$, $C_4H$, $C_4H_2$ were the new species produced but no C, $CH_3$, and $C_4H_4$ was observed. De Bleecker et al [21] has used Deschenaux's data to perform kinetic models and from these has predicted the formation of high mass hydrocarbon species at similar ratios to our $C_4H_2/C_2H_2$ ratio. Heavy species, reported in [21] have not been observed in our case and this may be due to the inhibition of the various polymerisation reactions by, for example, the presence of argon species which can lead to the charge exchange reaction [45].

$$Ar^+ + C_2H_2 \rightarrow C_2H_2^+ + Ar \qquad (13)$$

with a measured reaction rate coefficient $6 \times 10^{-18}$ m$^3$s$^{-1}$ [45].

In our case, neutrals species were also investigated at fixed pressure and varying bias. Neutrals other than primary gas inputs were detected by the mass analyser with the plasma off, and these can be attributed to production within the MEA. The production of $C_4H_2$ and $C_4H_4$ in the plasma indicates that carbon triple bonding remains mainly intact i.e. shows which supports the findings by Deschenaux [20] where spectra are bunched around even number carbon atoms. A potential reaction for the production of $C_4H_2$ in the plasma is the condensation reaction [21].

$$C_2H + C_2H_2 \rightarrow C_4H_2 + H \qquad (14)$$

However $C_4H_4$ production would require either a step-wise hydrogenation of $C_4H_2$ (possibly via $C_4H_3$), neutralisation of the $C_4H_4$ ion [23] or fragmentation of higher mass hydrocarbons etc.

### 4.2 CCP ionic species

The dominant ion species at 30 eV ion energy ($V_{dc}$ = 450 V, 6 mTorr) shown in figure 3 are $C_2H_2^+$ followed by $Ar^+$, $C_4H_2^+$ and $C_4H_3^+$. Energy integrated ion counts of the energy distribution shows the ranking to be $C_2H_2^+$ followed by $C_4H_2^+$ then $Ar^+$ and a later IED study confirmed this. These three species account for over 90% of all ions and possible production mechanisms for $C_4H_x^+$ family are given in equations (15 – 18). Deschenaux [20] observes significant species counts, however, from the heavy ion families up to $C_{15}H_y$. De Bleecker et al [21] has used Deschenaux's data to construct a suitable polymerisation model to account for these heavy ions. The presence of $Ar^+$ in our case is likely to disrupt or inhibit these polymerisation reactions [20, 38], the transparency of the plasma appears to be diminished. For ions, as with neutrals, we also note the carbon triple bonding remains mainly intact (even numbered carbon atom spectra), in agreement with [23].

$$C_2^+ + C_2H_2 \rightarrow C_4H^+ + H \qquad (15)$$
$$C_2H^+ + C_2H_2 \rightarrow C_4H_2^+ + H \qquad (16)$$
$$C_2H_2^+ + C_2H_2 \rightarrow C_4H_3^+ + H \qquad (17)$$
$$C_2H_3^+ + C_2H_2 \rightarrow C_4H_3^+ + H_2 \qquad (18)$$

### 4.3 ICP neutral species

The five dominant neutral species in the plasma are masses 30, 44, 18, 28 and 40 amu. Without bias, the Ar:$C_2H_2$ ratio is very large (~300), this falls to ~ 5-10 on application of both 8.311 MHz and 27 MHz bias, however it rises again under the high pressure and frequency condition. The greatest change in species counts upon plasma ignition is observed for masses 24 to 26 amu, reducing by two orders of magnitude, as noted above. Significant increases in masses 37 and 48 were also noted as well as the formation of new species at masses 68 and 69 amu. At the higher frequency, these heavier species are suppressed. Significant reduction in C and CH concentrations were observed as pressure was raised from 3.3 mTorr to 25 mTorr. $C_2$ is only observed in the plasma off state.

Hydrogen species (H, $H_2$ and $H_3$) were observed in small quantities; atomic hydrogen appearing close to the detection limit while $H_3$ was observed only at 27 MHz and $H_2$ was approximately constant across all conditions. Abstraction of hydrogen via electron dissociation of $C_2H_2$ [38], equation (1), should lead to significant atomic hydrogen however H is known to have a high wall sticking coefficient [32], an important factor in film formation. This high loss rate counteracts the production and the resultant H density is significantly lower than for low sticking efficiency. Further, its detection is affected, as in the case of $C_2H$ by interaction at the orifice. Formation of $H_2$ may also occur from electron dissociation of hydrocarbons [39]:

$$e + CH_2 \rightarrow C + H_2 + e \qquad (19)$$
$$e + C_2H_2 \rightarrow C_2 + H_2 + e \qquad (20)$$
$$e + C_2H_4 \rightarrow C_2H_2 + H_2 + e \qquad (21)$$
$$e + C_2H_4 \rightarrow C_2H + H_2 + H + e \qquad (22)$$
$$e + C_2H_6 \rightarrow C_2H_3 + H_2 + H + e \qquad (23)$$
$$e + C_2H_6 \rightarrow C_2H_2 + 2H_2 + e \qquad (24)$$

while the production of $H_3$, can occur via dissociation of $C_xH_{3+y}$. $H_3$ (and CH) are the only species that display a dependence on substrate frequency which may implicate sheath transport effects due to the lighter mass. This in turn may imply an ionic contribution to the $H_3$ production. Indeed $H_3^+$ is often the dominant ion in pure hydrogen plasmas [45]. Of course, the inevitable presence of $H_2O$ (mass 18) may be a more important factor in all observed $H_n$ counts.

The reaction paths for the formation of both C [39] and CH may follow equations (25 – 28) and the calculated rates are shown in Table 2

$$e + C_2H_2 \rightarrow C + CH_2 + e \qquad (25)$$
$$e + CH \rightarrow C + H + e \qquad (26)$$
$$CH_2 + H \rightarrow CH + H_2 \qquad (27)$$

$$e + C_2H_2 \rightarrow 2CH + e \qquad (28)$$

The presence of $C_2$ with no plasma is easily reconciled with production within the spectrometer as detailed above for $C_2H$. The lack of observed species once the plasma is ignited, is in accordance with the model predictions of De Bleecker et al [21].

As noted for the CCP case, the presence of heavy $C_xH_y$ (x = 4, 6, 8 etc.) is much less than observed by Deschenaux [20] and predicted by the model in [21]. In addition to the possible inhibiting effect of Ar, noted above, further $C_xH_y$ loss mechanisms include atomic H-induced fragmentation via

$$H + C_4H_3 \rightarrow C_2H_2 + C_2H_2 \qquad (29)$$
$$H + C_6H_3 \rightarrow C_4H_2 + C_2H_2 \qquad (30)$$

The rate coefficients for equations (29) and (30) are $1.1 \times 10^{-16}\,m^{-3}s^{-1}$ and $8.1 \times 10^{-17}\,m^{-3}s^{-1}$ respectively.
A significant (x 2 to x 6) increase in the ArH neutral is observed post ignition and is assumed to come from the ionic species which will be considered in the next section.

Decreases in $n_e$, $T_e$ and $V_{pl}$ were observed as acetylene was added to an argon plasma using the Langmuir probe, figure 10. One plausible cause for the observed decrease in electron density is polymerisation, eventually leading to the formation of nanometer scale particles, which act as a sink for electrons [46]. Another mechanism is electron attachment to neutrals giving rise to negative ions (which we have not investigated).

### 4.4 ICP ionic species

The dominant ionic species is $Ar^+$ even though the threshold ionisation energy of $C_2H_2$ [47] is lower than Ar [48]. The species with the next highest abundance is $ArH^+$. This is obviously a consequence of the loss of $C_2H_2$ neutrals. However, although we also observe $C_2H_2$ neutral loss mechanisms in the CCP, we do not observe the same $Ar^+$ dominance. From table 1, the ionisation threshold for $C_2H_2^+$ [47] (equation (8)) is similar to that of $Ar^+$ via the metastable pathway [49] equation (32). Thus, the reaction probabilities for both hydrogen extraction from $C_2H_2$ (equation (1)), typically requiring sufficient electrons of energies around 4 eV, and argon ionisation (equations. 31-33) must be lower in a CCP.

| Reactions | $E_{th}$ (eV) | Equation | Ref. |
|---|---|---|---|
| $e + Ar \rightarrow Ar^+ + 2e$ | 15.76 | 31 | [48] |
| $e + Ar \rightarrow Ar^m + e$ | 11.55 | 32 | [49] |
| $e + Ar^m \rightarrow Ar^+ + 2e$ | 4.21 | 33 | [49] |
| $e + C_2H_2 \rightarrow C_2H_2^+ + 2e$ | 11.41 | 8 | [47] |
| $e + C_2H_2 \rightarrow C_2H + H + e$ | 7.5 | 1 | [21] |

Table 1: Ionisation, metastable and dissociation threshold energies for reactions relevant to $Ar^+$ and $C_2H_2^+$ abundance.

ICPs are known to produce electron densities typically 1 – 2 orders of magnitude greater than CCPs for similar power inputs. Furthermore, electron energy distribution functions (EEDF) measured in the ICPs indicate a greater proportion of electrons of intermediate energy (3 eV- 11 eV) [11] compared to the bi-Maxwellian EEDFs expected in CCPs. This leads to greater $Ar^+$ densities in pure argon for ICPs. These conditions also favour reactions in equation (1) over equation (8), thus the greater loss of $C_2H_2$ in the ICP is expected. This explains the observed dependence in $Ar^+:C_2H_2^+$ ratio with plasma excitation mode. In fact, in the CCP, higher powers resulted in greater fall in pressure and hence $C_2H_2$, figure 8 & 9. Under the experimental conditions used here, the differences in plasma volume and the remote sampling position of the MEA in the CCP chamber would further contribute to this.

The production of $ArH^+$ ion has also been reported elsewhere [11, 50] and a number of production mechanism have been proposed [11]. All reactions involve an argon ion colliding with either a hydrocarbon molecule or hydrogen. In their ICP system with $CH_4/Ar$, Jie et al [11] observed an abundance of atomic H but they note that the argon ion attachment with atomic hydrogen has a low cross section [51]. The alternative reaction involving argon ions and hydrogen molecules, equation (34), is more plausible due to the observed $H_2$ count in our system, in agreement with that stated by Jie et al [11] and Rusu et al [50].

$$Ar^+ + H_2 \rightarrow ArH^+ + H \qquad (34)$$

This reaction is relatively fast (k ~ $1 \times 10^{-15}\,m^3\,s^{-1}$) [52], however the $H_2$ density is rather low for equation (34) to be the only source of $ArH^+$ production. A collection of other reactions are possible, involving hydrogen abstraction from $C_xH_y$

$$Ar^+ + C_xH_y \rightarrow C_xH_{y-1} + ArH^+ \qquad (35)$$

However the cross-sections for any of these reactions are not, to our knowledge, available.

The observed variation in ion flux with $V_{bias}$ is not consistent with the conventional Bohm theory [53] where flux depends only on ion density at the sheath ($n_i$) and electron temperature ($T_e$). The contribution of the rf bias field to local ionization is an obvious source of enhanced plasma density near the bias electrode. Further, the effect on local electron temperature may play a part. Any such changes however must be limited in extent since the Langmuir probe measurements (107 mm above the bias electrode) showed that the bulk plasma was only slightly affected by $V_{bias}$ up to 100 V. Methods of direct total ion flux measurements at the substrate are currently under investigation.

The only other report on $C_2H_2/Ar$ ionic species [24] was for an arc-filament Ti deposition plasma operated at ~ 10 mTorr where they observed $Ar^+/C_2H_2^+ \gg 1$. The dominant ion in their plasma was $C^+$ in agreement with our observation at high pressure $C_2H_2:Ar$.

A limited number of experiments were carried at much higher pressure (120 mTorr). Post-ignition, the pressure drop results in an argon-dominant plasma as before and $Ar^+$ is the dominant ion. The main carbon bearing ion species is $C^+$, figure 14. The threshold energy required for the formation of $C^+$ via dissociative ionization of $C_2H_2$, (equation (36) and plus similar), is near 20 eV [39, 54],

$$e + C_2H_2 \rightarrow C^+ + CH_2 + 2e \quad (36)$$

which is much greater than that for $C_2H_2^+$ (11.4 eV), $C_2H^+$ (16.5 eV) and $C_2^+$ (17.5 eV) [54]. Measurements in a similar system for pure argon indicated higher plasma density and lower electron temperature as the pressure is increased [55] [56]. The reduction in temperature should inhibit $C^+$ production while $Ar^+$ - based reactions, e.g. equation (37), need to be energetic and hence occur in the sheath [45].

$$Ar^+ + C_2H_2 \rightarrow C^+ + CH_2 + Ar \quad (37)$$

The presence of a much higher density of argon neutrals may play an important role, although possible mechanisms have not been investigated. The catalytic involvement of argon (ion or neutral) species, possibly at the chamber walls or bias electrode, therefore remains to be determined.

**4.5 Rates and Rate Coefficients**

In order to evaluate the various reaction possibilities we have considered the available rate coefficients and associated cross-sections [23, 27, 35, 42, 44, 45, 48, 49, 52, 57-60]. For the ICP electron collisions we have used our measured electron energy distribution functions (EEDFs) to determine the rate coefficients. The associated probability functions (EEPFs) are shown in figure 17, where $EEPF(E) = EEDF(E)/\sqrt{E}$. Analysis for pure Ar (thick dotted line) gives $n_e = 4.3 \times 10^{16}$ m$^{-3}$ and the expected [35, 61] combination of Maxwellian and Druyvesteyn populations. For the mixture, the fraction of final $C_2H_2$ is ~5% (post pressure drop) and the EEPF (thick solid line) shows a very similar form but with the number density $n_e$ reduced to $2.3 \times 10^{16}$ m$^{-3}$. A pure Maxwellian EEPF (thin straight line) is shown for comparison. In the absence of CCP measurements, we have assumed a bi-Maxwellian EEPF (dashed black line) with $n_e$ reduced by a factor of 30 compared with the $C_2H_2$:Ar ICP. Note that non-Maxwellian EEDFs are often characterized by an effective temperature < Te > [61]; detailed analysis of the EEDFs gave < Te > = 3.85 eV for all EEPFs shown in figure 17. The rate coefficients (k) and rates (r) for $C_2H_2$:Ar mixtures were calculated using the convolution of the energy-dependent EEDFs and cross-sections [39] and the outcomes are listed in Table 2. For comparison we also show f, the ratio of rate coefficient derived from measured (or assumed) EEDFs to those from the Maxwellian distribution with the same < Te >. We note that models (for pure $C_2H_2$) by Doyle J R [22] appear to be based on assumed ratios of rate coefficient while De Bleeker et al [21] and Bera et al [59] calculate EEDFs to determine their rates.

Rate coefficients for many of the heavy-heavy particle reactions mentioned in the text are also included in Table 2. The species densities used in these evaluations are either measured or estimated from mass spectra measurements. ICP reaction rate coefficients are generally about 20% less than those calculated using a pure Maxwellian, whilst the assumed CCP EEDF results in significantly higher rates. Of particular note is equation (14) which corresponds to the high abundance of $C_4H_2$ in figure

4. The rates also indicate that the main argon ionization route in the ICP is via metastables whereas in the CCP direct ionization dominates. In the ICP, the loss rates of $C_2H_2$ via equation (1) and (2) are significantly greater than through ionization. The measured abundances of $C_4H_x$ species (mass 49 to mass 53) matched the calculated rates for both CCP and ICP.

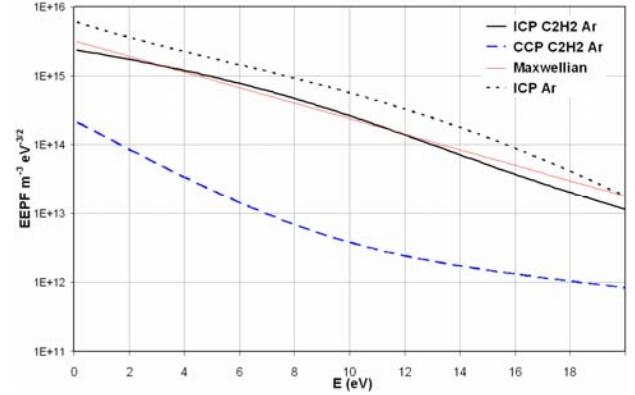

Figure 17: EEPFs: ICP, 200 W 13.56 MHz and $V_{bias}$ =0 V. Dotted line: 10 mTorr pure Ar, thick line: 3.3 mTorr $C_2H_2$:Ar (flow ratio 2:1), For comparison we also show a Maxwellian EEPF (thin straight line). This has the same electron density and temperature as that measured for the $C_2H_2$:Ar mixture in the ICP (thick line). The dashed shows the estimated EEPF for a $C_2H_2$:Ar CCP.

**5. Conclusions**

The neutral and positive ionic species measured in the CCP and ICP are the first such measurements published with $C_2H_2$:Ar as the working gas under bias and pressure conditions suited for diamond-like carbon deposition. In our companion paper [33] we investigate the ion energy distributions for each system and compare with the characteristics of deposited carbon films, under the same plasma conditions.

The drastic drop in chamber pressure at $C_2H_2$:Ar plasma turn on, can be explained in terms of the electron dissociation of $C_2H_2$, which leads to $C_2H$ formation. The $C_2H$ radical has a high surface loss probability factor [40] and is therefore lost to the chamber walls. It is thus the major component of film growth. The pressure drop along with $C_2H_2$ depletion, was most noticeable in the ICP because of the increased number density and the EEDF with sufficient electrons of energy E 11eV > E > 7.6 eV to lead to an argon dominated plasma irrespective of input gas ratio. The detection of $C_2H$ was inhibited due to high surface loss probability and a significant proportion of the observed $C_2H$ neutrals are due to the species generation within MEA. Depletion of the acetylene precursor was further investigated using infrared spectroscopy, where transient production of $CH_4$, showing a hitherto unreported $C_2H_2$-$CH_4$ relationship, was also observed. This transient $CH_4$ signal then decayed to almost zero at plasma equilibrium. The rise and fall time constants of $C_2H_2$ depletion were observed to be similar to those of the chamber pressure. Our spectra show heavier masses (> ~ 50 amu) are not as abundant as those reported in the literature for pure $C_2H_2$, suggesting argon inhibition of polymerization. However plasma transparency appeared to be diminished at higher pressures, possible evidence of nanoparticle suspensions; this was not observed at lower pressure. This study was focussed on positive ion and neutral species and negative ions were not investigated. An extensive set of rate coefficients and rates for the suggested electron-neutral and heavy-heavy particle reactions is presented using, where appropriate, realistic EEDFs and measured species densities. These are useful for incorporation in future models. These results have important implications for film growth studies and models, which are discussed in [33]. In particular, the argon species dominance in the ICP will result in an alternative ion bombardment mechanism to current models while the proposed high conversion rate to $C_2H$ is also critical since this radical is thought to be the dominant $sp^3$-promoting species during film formation.

| REACTION | Eqn. | | k (m$^{-3}$s$^{-1}$) | f | r (m$^{-3}$s$^{-1}$) | ref |
|---|---|---|---|---|---|---|
| $e + Ar \rightarrow Ar^+ + 2e$ | (31) | I | 4.5 x10$^{-16}$ | 0.66 | 5.5 x 10$^{19}$ | [48] |
| | | C | 3.6 x10$^{-15}$ | 5.16 | 1.5 x10$^{20}$ | |
| $e + C_2H_2 \rightarrow C_2H_2^+ + 2e$ | (8) | I | 1.0 x10$^{-15}$ | 0.67 | 3.0 x 10$^{16}$ | [39] |
| | | C | 6.0 x10$^{-15}$ | 3.82 | 2.6 x10$^{20}$ | |
| $e + Ar \rightarrow Ar^m + e$ | (32) | I | 1.6 x10$^{-16}$ | 0.71 | 2.0 x 10$^{19}$ | [48] |
| | | C | 5.4 x10$^{-16}$ | 2.35 | 2.3 x10$^{19}$ | |
| $e + Ar^m \rightarrow Ar^+ + 2e$ | (33) | I | 6.0 x10$^{-14}$ | 1.02 | 3.6 x10$^{20}$† | [48] |
| | | C | 4.7 x10$^{-14}$ | 0.80 | 2.3 x10$^{19}$† | |
| $e + C_2H_2 \rightarrow C_2H + H + e$ | (1) | I | 1.5 x10$^{-15}$ | 0.80 | 2.8 x 10$^{19}$ | [39] |
| | | C | 2.8 x10$^{-15}$ | 1.53 | 1.3 x10$^{20}$ | |
| $e + C_2H_2 \rightarrow$ all dissociation | (2) | I | 2.0 x10$^{-15}$ | 0.79 | 3.8 x 10$^{19}$ | [39] |
| | | C | 4.4 x10$^{-15}$ | 1.77 | 2.0 x10$^{20}$ | |
| $CH_2 + CH_3 \rightarrow C_2H_4 + H$ | (11) | I | 7 x10$^{-17}$ | | 3.1 x 10$^{19}$ | [44] |
| | | C | | | 3 x10$^{18}$ | |
| $e + C_2H_6 \rightarrow C_2H_4 + H_2 + e$ | (12) | I | 1.4 x10$^{-14}$ | 0.99 | 7.3 x 10$^{21}$ | [39] |
| | | C | 1.2 x10$^{-14}$ | 0.82 | 1.5 x10$^{18}$ | |
| $C_2H + C_2H_2 \rightarrow C_4H_2 + H$ | (14) | I | 5.8 x10$^{-17}$ | | 7.2 x 10$^{18}$ | [60] |
| | | C | | | 4.2 x10$^{22}$ | |
| $C_2^+ + C_2H_2 \rightarrow C_4H^+ + H$ | (15) | I | 1.7 x10$^{-15}$ | | 2.3 x 10$^{16}$ | [23] |
| | | C | | | 6.5 x 10$^{17}$ | |
| $C_2H^+ + C_2H_2 \rightarrow C_4H_2^+ + H$ | (16) | I | 1.7 x10$^{-15}$ | | 9.6 x 10$^{16}$ | [23] |
| | | C | | | 2.3 x10$^{18}$ | |
| $C_2H_2^+ + C_2H_2 \rightarrow C_4H_3^+ + H$ | (17) | I | 4.9 x10$^{-16}$ | | 2.7 x 10$^{17}$ | [23] |
| | | C | | | 7.7 x10$^{18}$ | |
| $C_2H_3^+ + C_2H_2 \rightarrow C_4H_3^+ + H_2$ | (18) | I | 7.2 x10$^{-16}$ | | 3.7 x 10$^{16}$ | [23] |
| | | C | | | 7.4 x 10$^{17}$ | |
| $Ar^+ + C_2H_2 \rightarrow C_2H_2^+ + Ar$ | (13) | C | 6 x10$^{-18}$ | | 8.0 x10$^{17}$* | [45] |
| $e + CH_2 \rightarrow C + H_2 + e$ | (19) | I | 2.3 x10$^{-16}$ | 0.83 | 3.9 x 10$^{18}$ | [39] |
| $e + C_2H_2 \rightarrow C_2 + H_2 + e$ | (20) | I | 2.8 x10$^{-16}$ | 0.75 | 5.4 x 10$^{18}$ | [39] |
| $e + C_2H_4 \rightarrow C_2H_2 + H_2 + e$ | (21) | I | 2.3 x10$^{-15}$ | 0.91 | 4.7 x 10$^{20}$ | [39] |
| $e + C_2H_4 \rightarrow C_2H + H_2 + H + e$ | (22) | I | 1.8 x10$^{-16}$ | 0.77 | 3.6 x 10$^{19}$ | [39] |
| $e + C_2H_6 \rightarrow C_2H_3 + H_2 + H + e$ | (23) | I | 6.5 x10$^{-17}$ | 0.73 | 3.3 x 10$^{19}$ | [39] |
| $e + C_2H_6 \rightarrow C_2H_2 + 2H_2 + e$ | (24) | I | 9.3 x10$^{-16}$ | 0.87 | 4.8 x 10$^{20}$ | [39] |
| $e + H_2 \rightarrow H + H + e$ | (38) | I | 1.7 x10$^{-15}$ | 0.79 | 1.2 x10$^{19}$‡ | [39] |
| $CH_2 + H \rightarrow CH + H_2$ | (25) | I | 2.7 x10$^{-16}$ | | 5.0 x 10$^{17}$ | [21] |
| $e + C_2H_2 \rightarrow C + CH_2 + e$ | (26) | I | 1.0 x10$^{-16}$ | 0.72 | 1.2 x 10$^{17}$ | [39] |
| $e + CH \rightarrow C + H + e$ | (27) | I | 4.5 x10$^{-15}$ | 0.89 | 9.8 x 10$^{18}$ | [39] |
| $e + C_2H_2 \rightarrow 2CH + e$ | (28) | I | 5.7 x10$^{-17}$ | 0.70 | 2.2 x 10$^{17}$ | [39] |
| $CH_2 + H_2 \rightarrow CH_3 + H$ | (4) | I | 5.2 x10$^{-17}$ | | 1.5 x 10$^{19}$ | [42] |
| $CH_3 + H_2 \rightarrow CH_4 + H$ | (3) | I | 6.9 x10$^{-20}$ | | 1.3 x 10$^{16}$ | [42] |
| $e + CH_4 \rightarrow CH_2 + H_2 + e$ | (39) | I | 5.1 x10$^{-16}$ | 0.81 | 2.8 x 10$^{19}$ | [39] |
| $Ar^+ + H_2 \rightarrow ArH^+ + H$ | (34) | I | 1.7 x10$^{-15}$ | | 3.9 x 10$^{18}$ | [52] |
| $e + C_2H_2 \rightarrow C^+ + CH_2 + 2e$ | (36) | I | 1.6 x10$^{-18}$ | 0.72 | 3.0 x 10$^{16}$ | [39] |
| $Ar^+ + C_2H_2 \rightarrow C^+ + CH_2 + Ar$ | (37) | I | 1 x10$^{-19}$ | | 6.3 x 0$^{14}$* | [45] |

Table 2: Rate coefficients k and rates r for selected reactions in the $C_2H_2$:Ar mixture ICP and CCP, f is the ratio of rate coefficient derived from realistic EEDFs to those from the Maxwellian distribution with the same $<T_e>$.
†The argon metastable ionization rates are calculated for $Ar^m$ densities of $3 \times 10^{17}$ m$^{-3}$ and $7 \times 10^{17}$ m$^{-3}$ for the ICP and CCP respectively.
\* These reactions only occur in the sheath as it requires fast ions.
‡ Underestimate since cross sections for all underlying processes are not available.